\documentclass[sigconf]{acmart}

\usepackage{multirow}
\usepackage{graphicx}
\usepackage{balance}
\usepackage{soul}
\usepackage{adjustbox}
\usepackage{amsmath}
\usepackage[utf8]{inputenc}
\usepackage{amsmath}
\usepackage{algorithm}
\usepackage{algorithmic}
\usepackage{float}

\AtBeginDocument{%
  }

\copyrightyear{2025} 
\acmYear{2025} 
\setcopyright{acmlicensed}\acmConference[SIGIR '25]{Proceedings of the 48th International ACM SIGIR Conference on Research and Development in Information Retrieval}{July 13--18, 2025}{Padua, Italy}
\acmBooktitle{Proceedings of the 48th International ACM SIGIR Conference on Research and Development in Information Retrieval (SIGIR '25), July 13--18, 2025, Padua, Italy}
\acmDOI{10.1145/3726302.3731943}
\acmISBN{979-8-4007-1592-1/2025/07}

\usepackage{caption}
\begin{document}

\title{IRA: Adaptive Interest-aware Representation and Alignment \\for Personalized Multi-interest Retrieval}

\author{Youngjune Lee}
\authornote{Both authors contributed equally to this research.}
\email{youngjune.lee93@navercorp.com}
\author{Haeyu Jeong}
\authornotemark[1]
\email{haeyu.jeong@navercorp.com}
\affiliation{%
  \institution{NAVER Corporation}
  \city{Seongnam}
  \country{Republic of Korea}
}

\author{Changgeon Lim}
\email{changgeon.lim@navercorp.com}
\author{Jeong Choi}
\email{jeong.choi@navercorp.com}
\author{Hongjun Lim}
\email{hongjun.lim@navercorp.com}
\affiliation{%
  \institution{NAVER Corporation}
  \city{Seongnam}
  \country{Republic of Korea}
}

\author{Hangon Kim}
\email{hangon.kim@navercorp.com}
\author{Jiyoon Kwon}
\email{jy.kwon@navercorp.com}
\author{Saehun Kim}
\email{saehun.kim@navercorp.com}
\affiliation{%
  \institution{NAVER Corporation}
  \city{Seongnam}
  \country{Republic of Korea}
}

\renewcommand{\shortauthors}{Youngjune Lee et al.}

\begin{abstract}

Online community platforms require dynamic personalized retrieval and recommendation that can continuously adapt to evolving user interests and new documents. However, optimizing models to handle such changes in real-time remains a major challenge in large-scale industrial settings. To address this, we propose the Interest-aware Representation and Alignment (IRA) framework, an efficient and scalable approach that dynamically adapts to new interactions through a cumulative structure. IRA leverages two key mechanisms: (1) Interest Units that capture diverse user interests as contextual texts, while reinforcing or fading over time through cumulative updates, and (2) a retrieval process that measures the relevance between Interest Units and documents based solely on semantic relationships, eliminating dependence on click signals to mitigate temporal biases. By integrating cumulative Interest Unit updates with the retrieval process, IRA continuously adapts to evolving user preferences, ensuring robust and fine-grained personalization without being constrained by past training distributions. We validate the effectiveness of IRA through extensive experiments on real-world datasets, including its deployment in the Home Section of NAVER’s CAFE, South Korea’s leading community platform.

\end{abstract}



\maketitle

\section{Introduction}
\begin{figure*}[t!]
  \centering
  \includegraphics[width=0.95\linewidth]{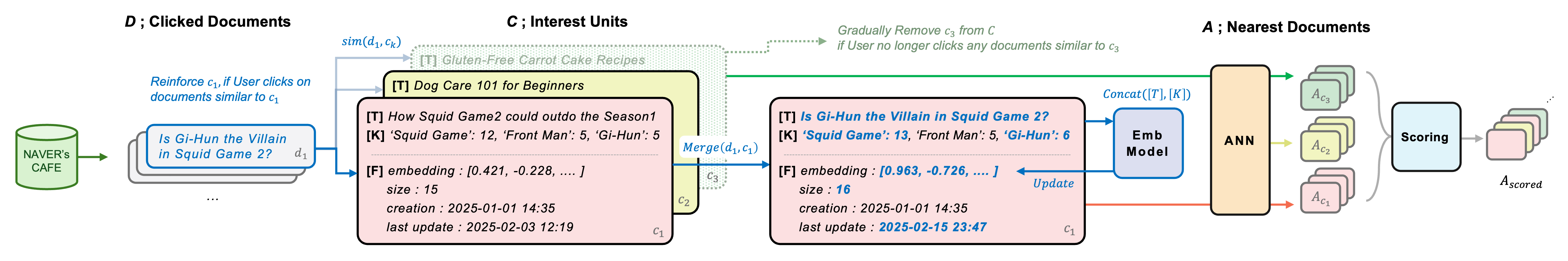}
  \caption{The overview of IRA pipeline. When the user clicks \(d_1\) similar to the existing Unit \(c_1\), \(c_1\) is reinforced. When the user no longer clicks any documents similar to \(c_3\), \(c_3\) is gradually removed. The original was in Korean, but translated to English.}
  \label{fig:overview}
\end{figure*}
Personalized retrieval and recommendation play a crucial role in various real-world applications, driving extensive research efforts to enhance their effectiveness \cite{sscdr, NeuMF, lightgcn, NGCF, bert4rec, sasrec, borisyuk2024lignn}. Widely used approaches typically encode user history as a sequence of item IDs and employ collaborative filtering \cite{rendle2012bpr, NeuMF} or sequential \cite{sasrec,bert4rec} models to generate user representations. However, collaborative filtering relies on item co-occurrence patterns, which struggle to fully capture multiple distinct interests within a single representation. Sequential models assume a smooth temporal evolution of interests, which limits their effectiveness for users with diverse and non-sequential preferences. Both approaches require frequent retraining to adapt to evolving behaviors and incorporate new items.

Feature-based \cite{dcn, guo2017deepfm, widedeep} and hybrid approaches \cite{s3rec, recformer, fdsa} mitigate some limitations by representing users and items as combinations of features, enabling more flexible modeling. However, they depend on domain-specific feature selection \cite{wang2022autofield, optfs, lee2023mvfs}, limiting adaptability across different scenarios. Moreover, their reliance on click-based interactions also makes them sensitive to patterns from specific time periods, requiring continuous model retraining.

Recent advances in Large Language Models (LLMs) \cite{llama, chatgpt, gpt4} have enabled robust personalization through textual representations \cite{bao2023tallrec, llmrec, collm}, effectively capturing user preferences. However, their slow inference and high computational costs make large-scale deployment impractical. These challenges underscore the need for a scalable framework that efficiently captures diverse user interests while dynamically adapting to user behavior.

To address these challenges, we propose the \textbf{Interest-aware Representation and Alignment (IRA)} framework, an efficient and scalable approach for personalized retrieval and recommendation. IRA dynamically captures evolving user interests through Interest Units, structured textual representations that encapsulate key aspects and recent interests of user interactions. Through cumulative updates, these Interest Units adaptively reflect both newly emerging and diminishing interests, while reinforcing persistent ones. IRA also leverages the semantic relationships between Interest Units and documents, captured by an embedding model fine-tuned to align them. This mitigates temporal biases in training data while eliminating the need for retraining, enabling IRA to seamlessly adapt to interest shifts and maintain high retrieval performance.

Through extensive experiments on real-world datasets, including an online A/B test conducted in the Home Section of NAVER’s CAFE\footnote{https://m.cafe.naver.com}, South Korea’s leading community platform, we demonstrate that IRA effectively captures diverse user interests.

\section{Methodology}
In this section, we describe our adaptive user interest modeling (Section \ref{Adaptive User Interest Modeling}), alignment between documents and user interests (Section \ref{Document Alignment}) and interest-aware retrieval process (Section \ref{Interest-aware Personalized Retrieval}). An overview of the entire pipeline is illustrated in Figure \ref{fig:overview}.

\subsection{Adaptive User Interest Modeling}\label{Adaptive User Interest Modeling}
We propose Interest Unit, which encodes user interests as contextual texts and dynamically adapts to user interactions. By ensuring that each Unit captures a distinct interest, IRA effectively represents multiple user preferences through a set of Units (Algorithm \ref{tab:context}).

Given a set of clicked documents $D$ = [$d_{1}, ..., d_{n}$], we generate a set of Units $C$ = [$c_{1}, ..., c_{k}$], where each $c_{*}$ contains a related set of documents $D_{c_{*}} \subset D$. Each Unit consists of (1) $[T]$, the title of the last clicked document in $D_{c_{*}}$, (2) $[K]$, key terms such as named entities extracted from titles within $D_{c_{*}}$ along with their occurrences, and (3) $[F]$, features such as the last update time of $c_{*}$ and the size of $D_{c_{*}}$. By forming the contextual text of $c_*$ through the concatenation of $[T]$ and $[K]$, our approach captures both recent and core interests while remaining inherently explainable.

When a user interacts with a new document $d$, its semantic similarity to existing Units is computed using the embedding model of Section \ref{Document Alignment}. If a relevant Unit $c' \in C$ that exceeds the threshold $\tau$ exists, $d$ is merged into $c'$. During the merging process: (1) $D_{c'}$ is updated to include $d$. (2) $[T]$ is updated to the title of $d$, and $[K]$ is aggregated with key terms extracted from $d$, with only the top-10 most frequent terms being used during inference. (3) The embedding of $c'$ is reconstructed using the updated $[T]$ and $[K]$. (4) $[F]$ is updated by summing numerical attributes or selecting the maximum. If no relevant Unit exists, a new Unit $c_{new}$ is created with $d$ as its initial element. When multiple relevant Units $C' \subset C$ exceed the threshold, the same process is applied, merging them into a single Unit $c_{merged}$.

To effectively optimize Interest Units over time, we adopt the pruning strategy for removing Units that are no longer interacted with. Considering that users dynamically consume both short-term and mid-to-long-term interests simultaneously, we design this strategy leveraging Unit features. For simplicity, we categorize Units into two groups: (1) $big$ (size $\geq$ 5), and (2) $small$ (size \textless 5). After updating a set of Units based on new interactions, we retain only the top 10 most recently updated Units from each group, ensuring that outdated Units gradually fade as new interests emerge.

Through a cycle of cumulative construction and pruning, each user's set of Units continuously adapts to new interactions, reinforcing frequently engaged Units while gradually pruning inactive ones that no longer receive clicks. This approach enables IRA to effectively capture users' diverse and evolving interests over time.

\begin{algorithm}[H]
\caption{Interest Unit Construction}
\label{tab:context}
\begin{algorithmic}[1]
\REQUIRE Clicked documents $D = \{d_1, d_2, \dots, d_N\}$, \\
         Current Interest Units $C = \{c_1, c_2, \dots, c_k\}$, \\
         Similarity function $\text{Sim}(d, c)$, Threshold $\tau$ for similarity

\FOR{$d \in D$}
    \STATE $C' \gets \{ c \in C \mid \operatorname{Sim}(d, c) \geq \tau \}$ \COMMENT{Find relevant Units}
    \IF{$C' \neq \emptyset$}
        \STATE $c_{\text{merged}} \gets \operatorname{Merge}(d, C')$ \COMMENT{Merge  into a single Unit}
        \STATE $C \gets (C \setminus C') \cup \{c_{\text{merged}}\}$ \COMMENT{Update Interest Units}
    \ELSE
        \STATE $c_{\text{new}} \gets \{d\}$ \COMMENT{Create new Unit}
        \STATE $C \gets C \cup \{c_{\text{new}}\}$
    \ENDIF
\ENDFOR

\RETURN $C$ \COMMENT{Updated Interest Units}
\end{algorithmic}
\end{algorithm}

\subsection{Document Alignment}\label{Document Alignment}
To ensure robust retrieval performance that remains unaffected by specific time periods or click patterns, we leverage the semantic relevance between Interest Units and documents. Since an Interest Unit consists of both the title and key terms, we tailored the embedding model to align with this structure, enabling the effective capture of both sentence-level semantics and keyword-level signals.

For training data construction, we randomly sampled search queries and retrieved 20 candidate documents for each query using in-house retrievers. Since not all retrieved documents are truly relevant, we leveraged a Korean specialized LLM \cite{yoo2024hyperclova} to classify them as either relevant or irrelevant with prompts from \cite{choi2024rradistill, rankgpt}. For negative sampling, we added two randomly selected documents from unrelated queries. Within the relevant set, we ranked documents based on the degree of key term overlap with the query, prioritizing those that share key terms to reinforce the model’s ability to capture keyword relevance effectively. For training, we fine-tune an in-house pre-trained Korean GPT \cite{gpt2} based embedding model (128M) using a combination of BCE and RankNet losses. This optimization allows the model to effectively capture the relevance between Interest Units and documents without being constrained by temporal biases introduced by training on click-based interactions.

\begin{algorithm}[H]
\caption{Interest-aware Personalized Retrieval}
\label{tab:ira}
\begin{algorithmic}[1]
\REQUIRE
    Units $C = \{c_1, c_2, \dots, c_k\}$,
    ANN function $\text{ANN}(c, N)$, 
    Similarity function $\text{Sim}(a, c)$
    
\STATE \textbf{Step 1: Document Retrieval by ANN}
\STATE $A \gets []$ \COMMENT{Aggregate ANN results from each Key Unit}
\FOR{$c \in C$}
    \STATE $A_{c} \gets \text{ANN}(\text{Emb}(c), N)$
    \STATE $A \gets A \cup A_{c}$
\ENDFOR
\STATE \textbf{Step 2: Scoring}
\FOR{$a \in A$}
    \STATE $a_{score} \gets \sum_{c \in C} \text{Sim}(a, c)$
\ENDFOR
\STATE $A_{\text{scored}} \gets \text{Sort}(A, \text{by } a_{\text{score}}, \text{descending})$
\end{algorithmic}
\end{algorithm}

\subsection{Interest-aware Personalized Retrieval}\label{Interest-aware Personalized Retrieval}
Leveraging our formulation of user interests as multiple contextual texts, IRA enables personalized document retrieval through semantic relevance. Given that users maintain multiple Units, we developed Interest-aware Personalized Retrieval, which integrates these diverse interests into the retrieval process (Algorithm \ref{tab:ira}).

Since both Interest Units and documents are embedded using the same model described in Section \ref{Document Alignment}, we can effectively apply an Approximate Nearest Neighbor (ANN) search to retrieve the top $N$ documents most relevant to each Unit. We then aggregate the documents $A_{c}$ for each Unit $c$ and produce the final results through the scoring step. To maintain efficiency without incorporating additional ranking models, each document's score $a_{score}$ is computed as the sum of its similarity to all $c \in C$. This score serves as the final ranking criterion, ensuring that the IRA framework retrieves documents that reflect users’ diverse interests simultaneously.

As a result, our approach ensures robust personalized retrieval while continuously adapting in near real-time to interest shifts, even in dynamic and large-scale environments.

\section{Experiment}
\label{sec:experiment}
\subsection{Setups}
\subsubsection{Dataset \& Evaluation}
We extracted one week of click logs from NAVER's CAFE, with dataset statistics in Table \ref{tab:dataset}. For a more effective evaluation of personalized recommendations, we randomly selected users with at least 15 clicks, while filtering out outliers with more than 200 clicks. To better reflect a continuous real-world recommendation setting, we designated each user's five most recent interactions as the test set and used the remaining data for Interest Unit construction and baseline training. Given the dynamic nature of the community platform with documents continuously being created, the test set includes a high proportion of cold items. To address this, we applied a cold item handling technique to all baselines, mapping unseen documents to their most semantically similar counterparts from the training set based on embedding similarity.

For evaluation, we employed widely used recommendation metrics \cite{NeuMF, derrd, sscdr}: Hit Ratio (H@N), NDCG (N@N). To mitigate the computational cost of large-scale user-item interactions, we followed the candidate sampling strategy used in \cite{transmetric, NeuMF, sscdr, dcd, derrd}. For this, we randomly selected 495 semantically distinct items per user based on low embedding similarity\footnote{We set a cosine similarity threshold of 0.4 to exclude highly similar titles.} to the test items. Each user was then evaluated five times, each time using a single test item from their test set and comparing it against the candidate set. The score per user was averaged over five evaluations, and the overall evaluation metric was computed as the mean score across all users.

\begin{table}[t!]
\caption{Data Statistics of NAVER's CAFE dataset.}
\label{tab:dataset}
\centering
\begin{adjustbox}{width=\linewidth}

\begin{tabular}{l|r|r|r|r}
\hline
\textbf{Dataset}           & \multicolumn{1}{c|}{\textbf{users}} & \multicolumn{1}{c|}{\textbf{items}} & \multicolumn{1}{c|}{\textbf{interactions}}  & \multicolumn{1}{c}{\textbf{cold items}} \\ \hline

\textbf{Train}           & 14,558                               & 248,075                                                                            & 659,283                                                                  & -                                \\
\textbf{Test} & 14,558                                  & 49,997                                                                             & 72,790                                                          & 19,149                          \\ \hline
\end{tabular}

\end{adjustbox}
\end{table}

\subsubsection{Baseline \& Implementation Details}
We compared IRA with representative recommendation models, each employing different optimization strategies: \textbf{ItemPop}, which ranks items by popularity; \textbf{MF-BPR} \cite{rendle2012bpr}, which optimizes matrix factorization using a pairwise ranking loss; \textbf{NeuMF} \cite{NeuMF}, which combines matrix factorization and MLP; \textbf{SASRec} \cite{sasrec}, which leverages attention mechanisms for sequential recommendation; and \textbf{Hybrid}, which integrates frozen text embeddings with trainable ID embeddings. While some hybrid approaches \cite{s3rec, recformer} exist, we simply concatenate document title embeddings with ID embeddings in SASRec because item attributes are not available in our setting. To generate embeddings for Units and documents, we utilized the model of Section \ref{Document Alignment}. For merging Units, we set the cosine similarity threshold $\tau$ to 0.65.

\subsection{Overall Performance}

Table \ref{tab:performance} presents overall offline performance. The comparison highlights differences in retrieving relevant documents and ranking effectiveness across various baselines. The results indicate that IRA successfully includes relevant documents in the top-n even with only pre-trained embedding model (no alignment), demonstrating its strong retrieval capability. Furthermore, the alignment process enhances both retrieval and ranking performance. This highlights IRA’s ability to comprehensively reflect users’ diverse interests.

\begin{table}[H]
\caption{Performance comparison. Best scores are in \textbf{bold}.}
\centering
\begin{adjustbox}{width=\linewidth}
\label{tab:performance}
\begin{tabular}{l|r|r|r|r|r|r}
\hline \hline
\multicolumn{1}{l|}{} & \multicolumn{1}{c|}{\textbf{H@5}} & \multicolumn{1}{c|}{\textbf{N@5}}  & \multicolumn{1}{c|}{\textbf{H@20}} & \multicolumn{1}{c|}{\textbf{N@20}}  &  \multicolumn{1}{c|}{\textbf{H@50}} & \multicolumn{1}{c|}{\textbf{N@50}} \\
\cline{1-7} 
\textbf{ItemPop}   & 0.0610 & 0.0366   & 0.1809 &  0.0701 & 0.3219 & 0.0979 \\
\textbf{MF-BPR}   & 0.4441& \textbf{0.3741}& 0.5551&  0.4062& 0.6330& 0.4216\\
\textbf{NeuMF}   & 0.4140& 0.2766&  0.5248&  0.3093&0.6007& 0.3243\\
\textbf{SASRec}   & 0.2527 & 0.1704    & 0.3885 &  0.2097  & 0.4951 & 0.2308\\
\textbf{Hybrid}   & 0.3962& 0.2612& 0.5860&  0.3165& 0.7018& 0.3396\\\cline{1-7}
\textbf{IRA (no alignment)}    & 0.4340 & 0.2860  & 0.6092 &  0.3372& 0.7214 & 0.3595 \\
\textbf{IRA (ours)}   & \textbf{0.5687} & 0.3677   & \textbf{0.7043} &  \textbf{0.4074}  & \textbf{0.7862}& \textbf{0.4237} \\
\hline \hline
\end{tabular}
\end{adjustbox}
\end{table}

\subsection{Study of IRA}

To further investigate the effectiveness of IRA, we conducted various experiments using three datasets over consecutive weeks: the initial training period (A), followed by two consecutive weeks denoted as (B) and (C). For evaluation, period (C) was refined to include only each user's last five clicks.

\subsubsection{Impact of the number of Units} 
\begin{figure}[H]
    \centering
    \begin{minipage}{0.49\linewidth}
        \centering
        \includegraphics[width=\linewidth]{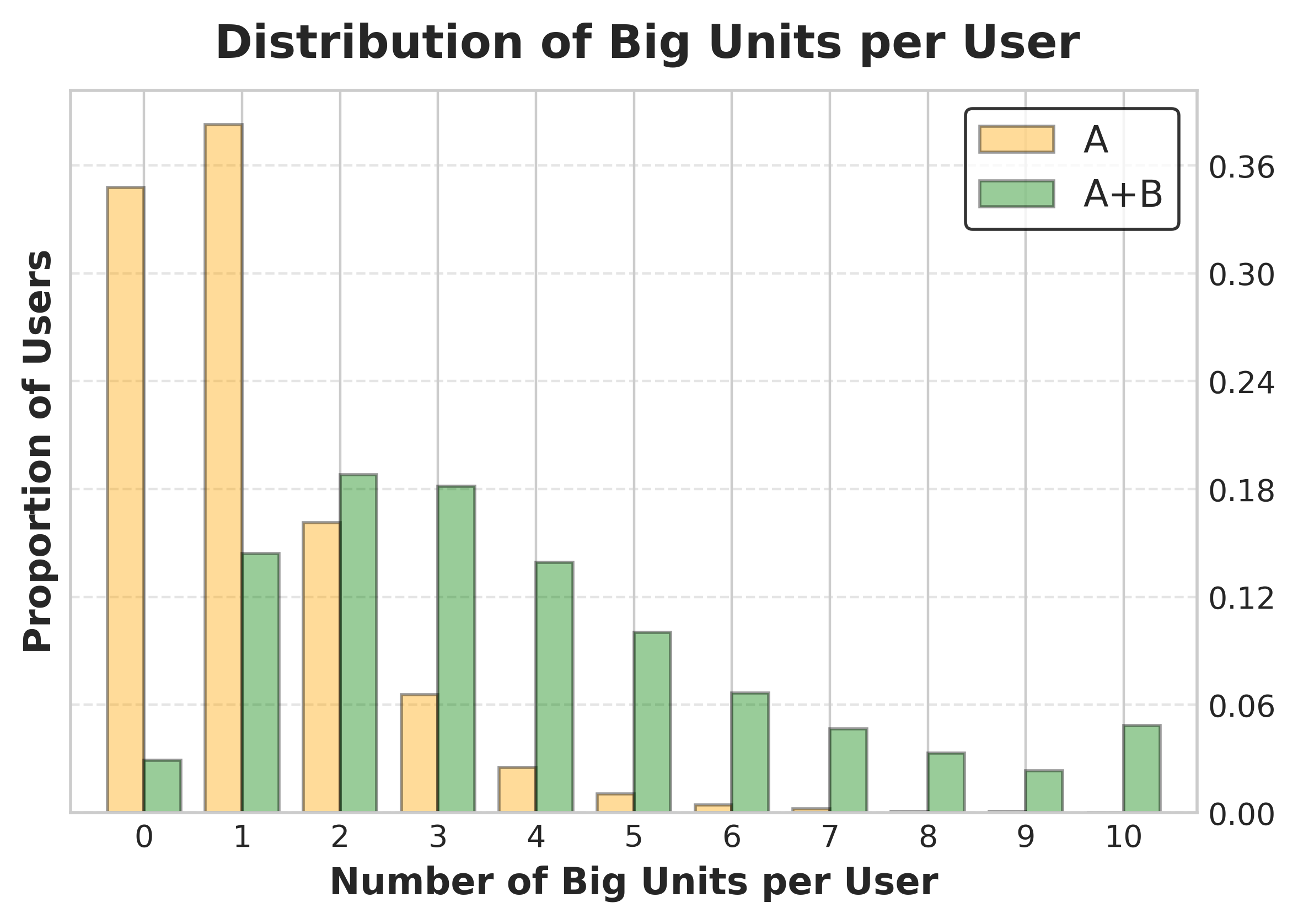}
    \end{minipage}
    \begin{minipage}{0.49\linewidth}
        \centering
        \includegraphics[width=\linewidth]{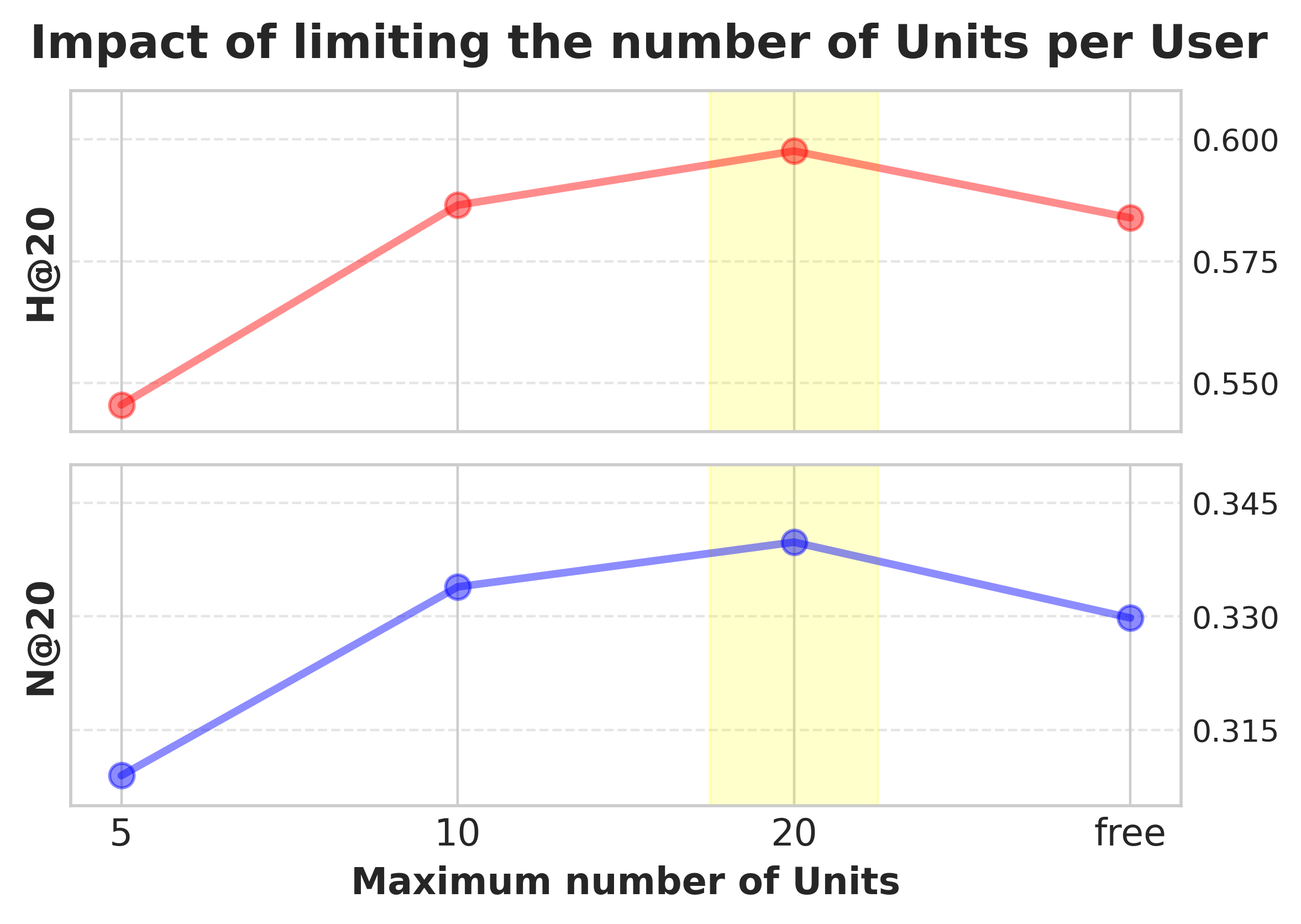}
    \end{minipage}
    \caption{Analysis of big Unit distribution (Left) and the number of Units (Right). }
    \label{fig:fish}
\end{figure}

We analyzed how the number of big Units per user changed from period A to A+B. As shown in Figure \ref{fig:fish} (Left), most users initially had only one or two big Units. However, after incorporating period B, the majority possessed a significantly larger number. This indicates that users gradually diversify their interests and engage more deeply over time.

To examine the impact of adaptive Unit construction on modeling evolving interests, we evaluated the impact of limiting the number of Units per user during period A+B, setting the maximum to 5, 10, 20, or unconstrained (free). As shown in Figure \ref{fig:fish} (Right), performance declines when Unit count is either too small or left unconstrained. This highlights the necessity of IRA's dynamic adaptation mechanism, as users simultaneously explore diverse interests while some are no longer preferred and gradually become inactive.

\subsubsection{Adaptability of interest shifts}

To examine the temporal robustness of our approach, we evaluated the performance on period C without retraining (MF-BPR, NeuMF, Hybrid-A, IRA-A). As shown in Table \ref{tab:adapt}, all baselines exhibited a notable decline in performance compared to IRA, which maintained relatively strong results, demonstrating its temporal robustness.

To further evaluate the adaptability to interest shifts,  we evaluated models incorporating interactions from period B. As Hybrid is designed as a sequential model, we allowed it to utilize period A+B for inference (Hybrid A+B). The results indicated that, despite leveraging additional signals from period B, sequential approach exhibited no significant improvements. In contrast, IRA showed consistent gains when incorporating period B into Interest Unit construction (IRA A+B), demonstrating its adaptability to shifting user interests while preserving robustness.

\begin{table}[t!]
\caption{Adaptability study of Interest Shifts.}
\centering
\begin{adjustbox}{width=\linewidth}
\label{tab:adapt}
\begin{tabular}{l|r|r|r|r|r|r}
\hline \hline
\multicolumn{1}{l|}{}     & \multicolumn{1}{c|}{\textbf{H@5}}   & \multicolumn{1}{c|}{\textbf{N@5}} & \multicolumn{1}{c|}{\textbf{H@20}} & \multicolumn{1}{c|}{\textbf{N@20}} & \multicolumn{1}{c|}{\textbf{H@50}} & \multicolumn{1}{c}{\textbf{N@50}} \\
\cline{1-7} 
\textbf{MF-BPR}    & 0.1875 & 0.1340 & 0.2921  & 0.1638  & 0.3872 & 0.1826  \\
\textbf{NeuMF}    & 0.1202 & 0.0903 & 0.1766  & 0.1061  & 0.2501 & 0.1205  \\ 
\textbf{Hybrid (A)}    & 0.2199 & 0.1515 & 0.3531  & 0.1895  & 0.4720 & 0.2130  \\ 
\textbf{Hybrid (A+B)}    & 0.2248 & 0.1527 & 0.3543  & 0.1897  & 0.4698 & 0.2125  \\ \cline{1-7} 
\textbf{IRA (A)}    & 0.4366 & 0.2857 & 0.5693  & 0.3242  & 0.6646 & 0.3431  \\ 
\textbf{IRA (A+B)}    & 0.4583 & 0.2993 & 0.5976  & 0.3398  & 0.6948 & 0.3591  \\ 
\hline \hline
\end{tabular}

\end{adjustbox}
\end{table}

\subsubsection{Impact of Pruning Strategies}
\begin{figure}[H]
    \centering
    \begin{minipage}{\linewidth}
        \centering
        \includegraphics[width=\linewidth]{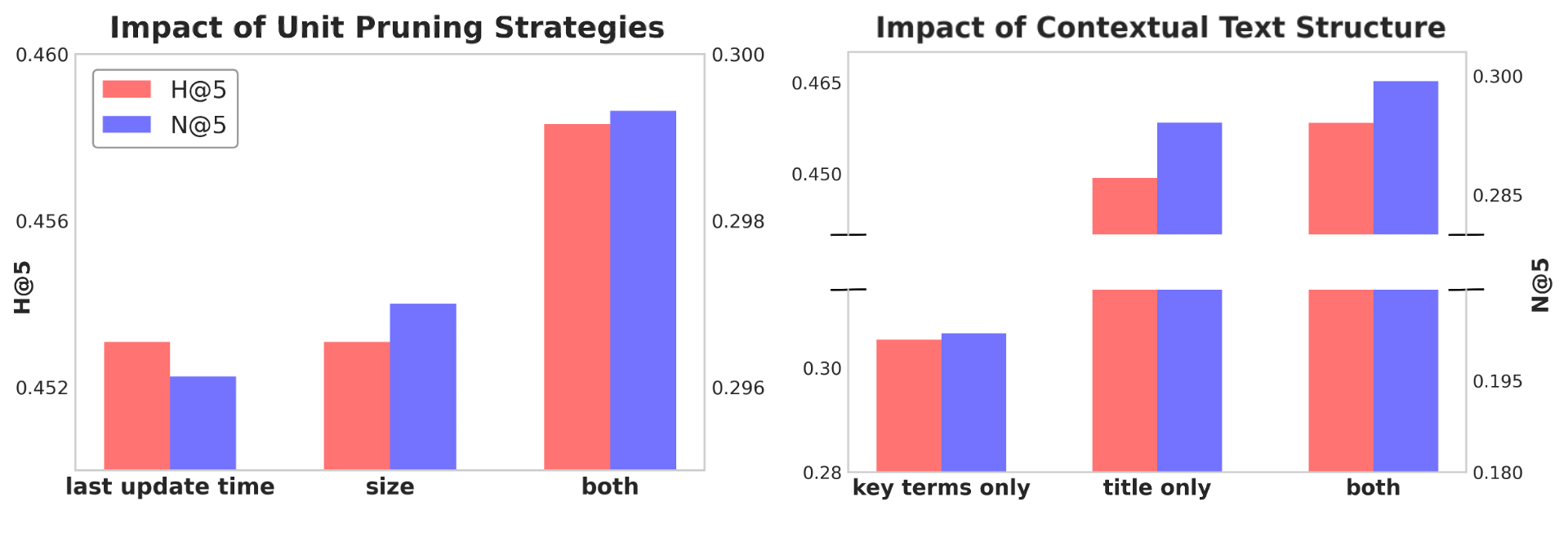}
    \end{minipage}
    \caption{Analysis of Unit pruning strategies (Left) and contextual text construction (Right). }
    \label{fig:ablation}
\end{figure}

To assess the impact of different pruning strategies, we tested two alternative approaches: (1) retaining the 20 most recently updated Units (last update time) without considering size, and (2) retaining the 20 largest Units (size) without considering recency. As shown in Figure \ref{fig:ablation} (Left), both strategies resulted in lower performance compared to incorporating both factors. This highlights the importance of balancing both the intensity and recency of user interests when refining Units, ensuring that significant interests are retained effectively.

\subsubsection{Contextual text structure}
To assess the impact of contextual text structure on performance, we conducted an ablation study by evaluating three variations: (1) using only key terms, (2) using only the title, and (3) using both. As shown in Figure \ref{fig:ablation} (Right), using only key terms or only title results in lower performance compared to using both key terms and the title together. These results highlight the importance of incorporating the title of the last clicked document, as it effectively captures a user's latest interests, while key terms provide the core aspects of interest and further enhance performance by complementing the title.

\subsubsection{Online A/B test}

We integrated IRA’s personalized retrieval into the Home Section of NAVER's CAFE, which previously featured only generally popular and explicitly favorited channel contents. We then conducted an online A/B test for two weeks. As a result, time spent per document and the total number of clicks increased by 1.2\% and 5.4\%, respectively, while overall user engagement time across the entire CAFE service grew by 1\%. These results demonstrate that IRA effectively aligns with users' diverse interests, enhancing user satisfaction and improving overall engagement.

\section{Conclusion \& Future Work}
We propose IRA, an efficient and scalable framework that dynamically adapts to users' evolving interests through the cumulative approach. By leveraging Interest Units and the retrieval process, IRA achieves robust performance in dynamic real-world environments. Through extensive experiments and analysis, we demonstrate the effectiveness of IRA and its successful deployment in the Home Section of NAVER's CAFE platform. Our results highlight IRA’s strong performance in large-scale industrial settings, reinforcing its practicality for real-world personalized retrieval.

In future work, we will further explore IRA’s integration with other retrieval methods and leverage its flexible design to enhance user interest modeling beyond direct interactions, such as incorporating Interest Units from users with similar interest patterns as a collaborative signal. We believe that our approach to dynamically adapting to evolving multi-interest user behaviors provides valuable direction for practical implementations for real-world applications.

\section{Presenter Bio}
\textbf{Youngjune Lee} is a machine learning engineer at NAVER. He earned his Master’s degree from KAIST, South Korea. His research focuses on personalization, retrieval, and ranking models.\\
\textbf{Haeyu Jeong} is a machine learning engineer at NAVER. Her research interests are in personalization, information retrieval, and user modeling.

\bibliographystyle{ACM-Reference-Format}
\balance
\bibliography{custom}

\end{document}